\documentclass[aps,prd,twoside,onecolumn,superscriptaddress,floatfix,nofootinbib]{revtex4}
\pdfoutput=1
\usepackage{amsmath}
\usepackage{bm}
\usepackage{xcolor}

\newcommand\beq{\begin{equation}}
\newcommand\eeq{\end{equation}}
\newcommand\beqn{\begin{eqnarray}}
\newcommand\eeqn{\end{eqnarray}}
\newcommand\lsim{\mathrel{\rlap{\lower4pt\hbox{\hskip1pt$\sim$}}
        \raise1pt\hbox{$<$}}}
\newcommand\gsim{\mathrel{\rlap{\lower4pt\hbox{\hskip1pt$\sim$}}
        \raise1pt\hbox{$>$}}}

\newcommand{\aaps}{{Astron.~Astrophys.~Supp.}}

\newcommand{\aap}{{Astron.~Astrophys.}}

\newcommand{\apjs}{{Astrophys.~J.~Supp.}}
\newcommand{\aj}{{Astron.~J.}}
\newcommand{\mnras}{{Mon.~Not.~R.~Astron.~Soc.}}
\usepackage{graphicx}
\usepackage[large]{subfigure}
\usepackage{amssymb, amsmath}
\usepackage{natbib}
\newcommand{\ie}{\emph{i.e.}}

\newcommand{\Planck}{\emph{Planck}}
\newcommand{\WMAP}{\emph{WMAP}}
\newcommand{\Bicep}{{\sc Bicep2}}
\newcommand{\B}{{\sc Bicep1}}

\begin{document}

\title{Toward an Understanding of Foreground Emission in the \Bicep\  Region}

\author{Raphael~Flauger}
\affiliation{Institute for Advanced Study, Einstein Drive, Princeton, NJ 08540, USA}
\affiliation{CCPP, New York University, New York, NY 10003, USA}

\author{J.~Colin~Hill}
\affiliation{Dept.~of Astrophysical Sciences, Peyton Hall, Princeton
  University, Princeton, NJ 08544, USA}
\author{David~N.~Spergel}
\affiliation{Dept.~of Astrophysical Sciences, Peyton Hall, Princeton
  University, Princeton, NJ 08544, USA}

\begin{abstract}
\Bicep\  has reported the detection of a degree-scale $B$-mode
polarization pattern in the Cosmic Microwave Background (CMB) and has
interpreted the measurement as evidence for primordial gravitational
waves. Motivated by the profound importance of the discovery of
gravitational waves from the early Universe, we examine to what extent
a combination of Galactic foregrounds and lensed $E$-modes could be
responsible for the signal. 
We reanalyze the \Bicep\ results and show that the 100$\times$150 GHz and
150$\times$150 GHz data are consistent with a cosmology with $r=0.2$ and negligible foregrounds,
but also with a cosmology with $r=0$ and a significant dust polarization signal.
We give independent estimates of the dust polarization signal in the
\Bicep\  region using a number of different approaches: (1) data-driven models
based on \emph{Planck} 353~GHz intensity, polarization fractions
inferred from the same \emph{Planck} data used by the \Bicep\  team but corrected for
CMB and CIB contributions, and polarization angles from starlight
polarization data or the Planck sky model; (2) the same set of pre-\Planck\
models used by the \Bicep\ team but taking into account the higher polarization fractions
observed in the CMB- and CIB-corrected map;
(3) a measurement of neutral hydrogen gas column density $N_{\rm H\textsc{i}}$ in the
\Bicep\  region combined with an extrapolation of a relation between
H{\sc i} column density and dust polarization derived by \Planck;
and (4) a dust polarization map based on digitized \Planck\ data,
which we only use as a final cross-check.   While these approaches are
consistent with each other,  the expected amplitude of the
  dust polarization power spectrum remains uncertain by about a factor of
  three. The lower end of the prediction leaves room for a primordial
contribution, but at the higher end the dust in combination with the
standard CMB lensing signal could account for the \Bicep\  observations,
without requiring the existence of primordial gravitational waves.
By measuring the cross-correlations between the pre-\Planck\ templates
used in the \Bicep\ analysis
and between different versions of a data-based template, we emphasize that
cross-correlations between models are very sensitive to noise in the
polarization angles and that measured cross-correlations are likely underestimates of the 
contribution of foregrounds to the map. These results suggest  that \B\ and \Bicep\  data alone
  cannot distinguish between foregrounds and a primordial gravitational wave~signal, and that future Keck Array observations at 100~GHz and \Planck\ observations
at higher frequencies will be crucial to determine whether the signal is of primordial origin.  
\end{abstract}

\maketitle


\section{Introduction}

The \Bicep\  collaboration has made the deepest map of the microwave sky and detected degree-scale $B$-mode polarization fluctuations in the Cosmic Microwave Background (CMB)~\cite{B2,B2data}.  If these fluctuations are sourced by cosmological tensor modes, the implications for fundamental physics would be profound~\cite{Starobinsky:1979ty,polnarev85,Davis1992,Crittenden1993,Seljak1997,Seljak-Zaldarriaga1997,Kamionkowski1997}.  The signal would constitute direct evidence for quantum fluctuations in the spacetime metric. It would provide very strong additional support that a period of cosmic inflation occurred in the early Universe~\cite{Starobinsky:1979ty,Guth1981,Linde1982,Albrecht-Steinhardt1982}.  The inferred amplitude of the tensor modes would provide a measurement of the Hubble rate during inflation ($H\approx 10^{14}$ GeV)~\cite{Starobinsky:1979ty} and evidence that the inflaton underwent a trans-Planckian excursion in field space~\cite{Lyth1997}. In addition, it would have important implications for axion physics~\cite{Fox:2004kb,Marsh:2014qoa}, would essentially exclude cosmologically stable moduli with masses below \mbox{$10^{14}$ GeV}, and would motivate serious consideration of the gravitino problem. It would also place a direct bound on the graviton mass, $m_{g}\!\lsim 3 \times 10^{-28}$~eV~\cite{Dubovsky:2009xk}. These implications would hold regardless of the precise value of the tensor-to-scalar ratio. The relatively large amplitude of the tensor modes implied by the \Bicep\  signal ($r \approx 0.2$) would promise percent-level detections in coming years, which would be extremely valuable to distinguish between different inflationary models and reheating scenarios. It would also allow for the possibility of measuring the tensor tilt and testing the inflationary consistency condition.

However, the foreground estimates presented in the \Bicep\ analysis already show that an understanding of foregrounds is critical for the cosmological interpretation. While the data imply a tensor-to-scalar ratio of $r=0.2^{+0.07}_{-0.05}$ if foregrounds are not subtracted, one infers $r=0.12^{+0.05}_{-0.04}$ after subtracting a level of foregrounds given by the auto-power spectrum of the ``DDM2'' foreground model in~\cite{B2}. For the same model, the preliminary \Bicep\ $\times$ Keck Array data would indicate a tensor-to-scalar ratio as small as $r=0.06^{+0.04}_{-0.03}$.

Motivated by the importance of a  detection of primordial
gravitational waves, this paper aims to make progress on the
characterization of foreground emission in the region of sky observed
by the \Bicep\  collaboration.  At \mbox{150 GHz}, polarized dust
emission is expected to be the dominant polarized foreground, while
polarized synchrotron emission is expected to dominate at lower
frequencies.

The \WMAP\ K (23 GHz) and Ka (33 GHz) band polarization maps provide useful templates for estimating the synchrotron contribution~\cite{Bennettetal2013}.  The \Bicep\  analysis used the auto-spectrum of the K-band map and its cross-spectrum with the \Bicep\  map to set an effective upper limit of $r_{\rm synch}\!< 0.003$ on synchrotron contamination, assuming a spectral index of $\beta_{\rm synch} = -3.3$.  This may underestimate the importance of synchrotron emission, as the spectral index in the \Bicep\  region is rather uncertain~\cite{Fuskeland2014}.  Ref.~\citep{Bennettetal2013} finds $\beta_{\rm synch} \approx -3.1$ at the lowest frequency range.  Cross-correlating the \WMAP\ measurements, \Planck\ LFI observations, and the \Planck\ HFI measurements, Ref.~\citep{Planck2014dustfreq} reports a spectral index of $\beta_{\rm synch}=-2.92 \pm 0.02$ for dust-correlated synchrotron over 39\% of the sky. Radio loops might cause a significant departure from this value of the spectral index~\cite{Wolleben:2007pq,Liu:2014mpa}, but we have found no evidence for this in the \Bicep\ region of the sky. Assuming a spectral index of $\beta_{\rm synch}=-2.92$ to extrapolate to 150 GHz implies that synchrotron radiation would account for  10\,--\,15\% of the amplitude of the signal in the map at 150 GHz, important enough to include in our analysis below but almost certainly not large enough to account for the observed~signal. 

Dust polarization is potentially a more significant foreground.  
Although the \Bicep\  region is low in Galactic dust emission and gas
column density, the region appears to be more polarized than average.
The three-year \WMAP\ K-band data already revealed that the
large-scale synchrotron polarization fraction in this region is
$\sim\!30\%$ (see {\it e.g.}  Fig.~4 in Ref.~\cite{Kogut:2007tq}),
and \Planck\ has shown that the \Bicep\  region overlaps the $30$\%,
$40$\%, and $50$\% contours in an all-sky thresholded map of polarized
intensity~\cite{AumontESLAB}. The very low dust intensity in
the \Bicep\  region therefore does not  guarantee negligible
polarized dust~emission.

This paper is organized as follows.  In
section~\ref{sec:multi_fit}, we discuss the prediction of the null
hypothesis of a combination of synchrotron emission, dust emission,
and lensed $E$-modes, and show that current {\sc Bicep1} and \Bicep\  data cannot convincingly rule out the null
hypothesis based on a measurement of the spectral index.  
Fortunately, the \Planck\ satellite provides a powerful tool for
studying dust polarization with its 217 and 353 GHz sky
maps~\cite{Planck2014dustoverview}. A detailed characterization of
the dust emission in the \Bicep\ region 
awaits the public release of the data and their analysis by the
\Planck\ collaboration, but we attempt to make progress toward the
understanding of foregrounds in the \Bicep\ region in section~\ref{sec:dust} by using temperature
and polarization data from \WMAP, temperature data from \Planck, early
\Planck\ polarization results shown at the ESLAB meeting in April
2013\footnote{{\tt
    http://www.rssd.esa.int/index.php?project=PLANCK\&page=47\char`_eslab
}}, starlight polarization data, and new \Planck\ polarization results
presented in a recent series of
papers~\cite{Planck2014dustfreq,Planck2014dustoverview,Planck2014duststar}.
According to our estimates, the amount of polarized dust emission is uncertain and could
potentially be large enough to account for the excess $B$-mode
power seen by \Bicep. To understand to what extent the cross-spectra
presented by the \Bicep\  collaboration provide convincing evidence that
foregrounds are subdominant, we compute cross-spectra for a suite of
ninety-six foreground models in section~\ref{sec:x-corr} as well as
the pre-\Planck\ foreground models considered by \Bicep. We show that
the noise in polarization angles in templates can lead to significant
underestimates of the foreground contribution.   Our
conclusions are presented in section~\ref{sec:outlook}.


\section{Testing the Null Hypothesis}
\label{sec:multi_fit}
\begin{figure}
\includegraphics[trim=.05cm 0cm .5cm .4cm,clip,width=3.4in]{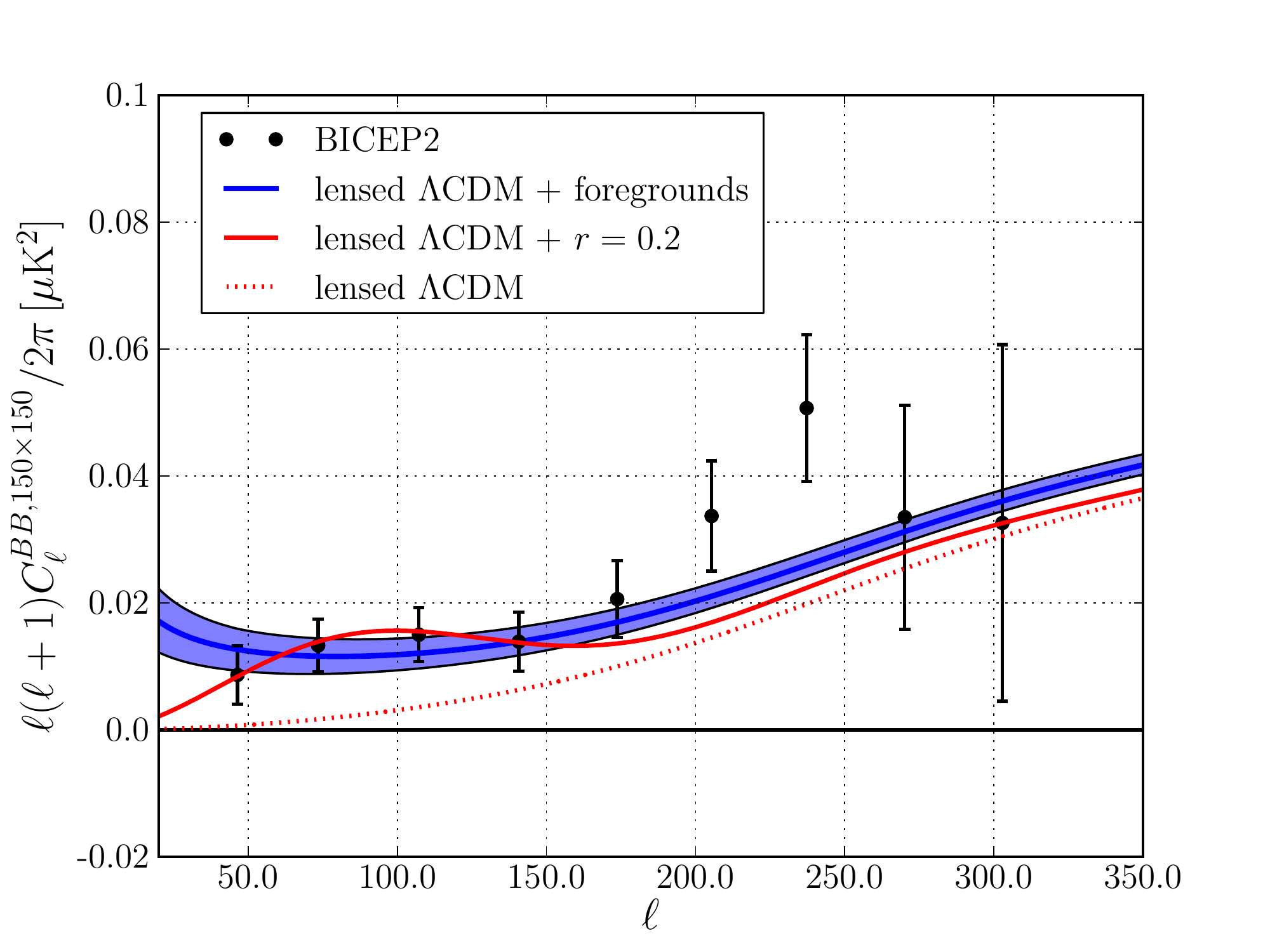}
\includegraphics[trim=.05cm 0cm .5cm .4cm,clip,width=3.4in]{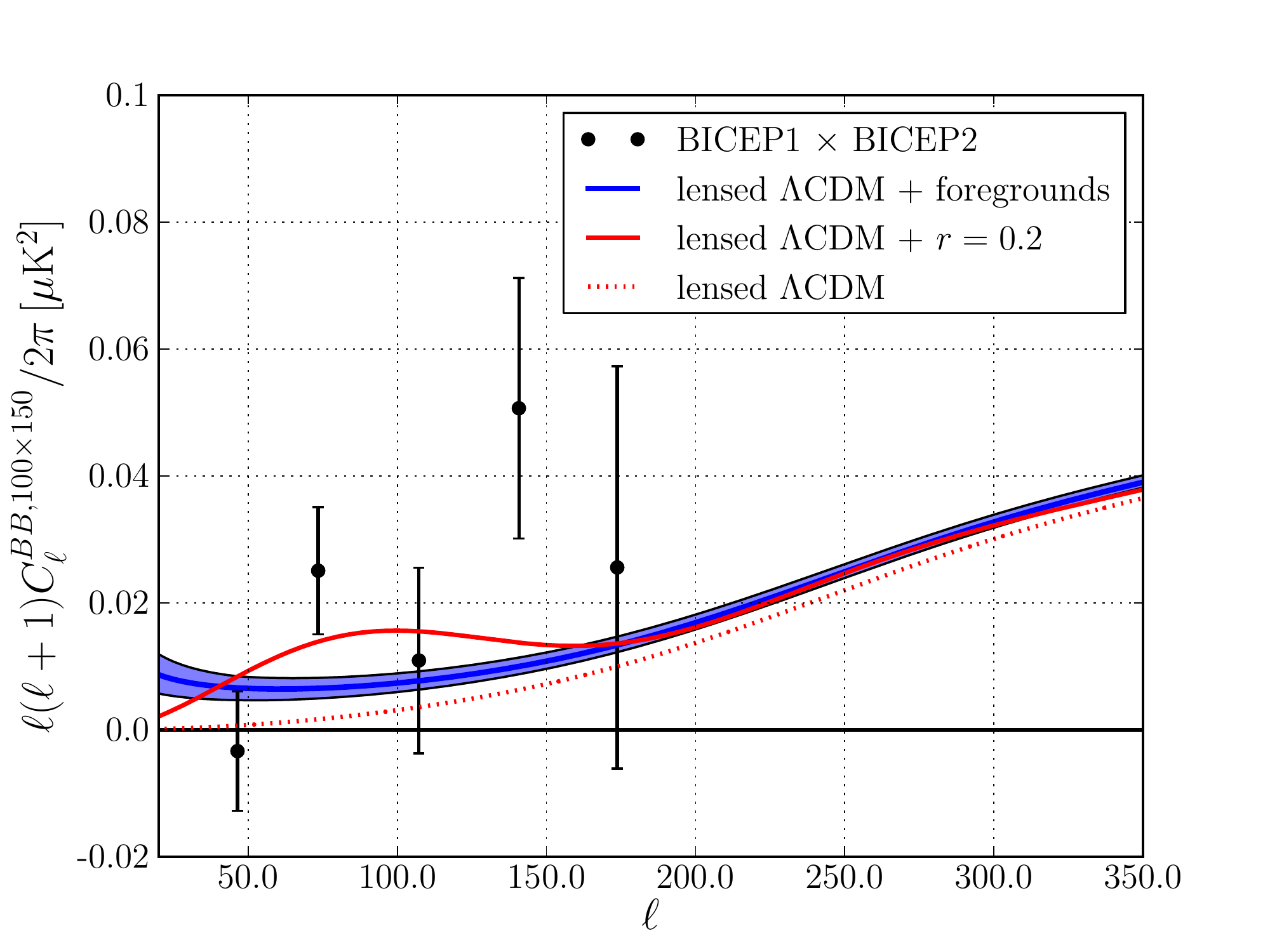}
\caption{The left panel compares the
  \Bicep\ $\times$ \Bicep\ signal to two models: the best-fit lensed
  CMB plus gravitational wave model and the best-fit lensed CMB plus
  Galactic foregrounds model.  The right panel compares the
  \B\ $\times$ \Bicep\ signal to the same set of models with the same
  parameters.  The shaded band (light blue) includes uncertainties on the amplitude of the dust
  obtained from the \Bicep\ $\times$ \Bicep\ fit, as well as uncertainties on the
  synchrotron amplitude and scaling with $\ell$. The black error bars in both panels include sample variance for a $\Lambda$CDM cosmology with $r=0.2$.
 \label{fig:spectra}}
 \end{figure}
 
In this section, we assess whether the 100 and 150 GHz \B\ and
\Bicep\ data alone can discriminate between a cosmological signal and
the null hypothesis of a combination of synchrotron, dust, and lensed
$E$-modes.  This question was also addressed by Mortonson
and Seljak~\cite{Mortonson2014}, who reached similar
conclusions.

We define the null
hypothesis as a $\Lambda$CDM cosmology with
$r=0$, polarized synchrotron with a level estimated  from \WMAP\ data, and polarized dust emission
of unknown amplitude.
For synchrotron, we measure the \WMAP\ K-band polarization power
spectrum on the \Bicep\ patch and extrapolate to shorter wavelengths
assuming a spectral index of $\beta_{\rm synch} = -2.92$, which corresponds to the
mean index over 39\% of the sky recently derived
by~\cite{Planck2014dustfreq}.  At $\ell = 46$, we find
$\ell(\ell+1)C_\ell^{BB}/2\pi=(4.0 \pm 2.7)\times 10^{-4} \, \mu{\rm K}^2$
and $(6.9 \pm 4.6)\times 10^{-5} \, \mu{\rm K}^2$ at 100 and 150~GHz,
respectively, after filtering the maps using the \B\ transfer
function~\cite{BarkatsBICEP}.  
We assume that the synchrotron power spectrum scales as
$C_{\ell}^{BB} \propto \ell^{-2.6}$, consistent with
\WMAP\ measurements~\cite{Page07} and radio maps~\cite{LaPorta08}.
Based on the cross-correlation between starlight dust polarization and
Galactic synchrotron directions~\cite{Page07} and the results presented in~\cite{Planck2014dustfreq}, we assume that the
polarized dust and synchrotron emissions are 70\% correlated.   Since
the amplitude of the dust polarization signal is unknown, we allow it
to vary and assume that the dust power spectrum scales as
$C_{\ell}^{BB} \propto \ell^{-2.4}$~\cite{AumontESLAB}.  We find that the
best-fit amplitude for the polarized dust component in the
\Bicep\ 150~GHz data is $\ell(\ell+1)C_\ell^{BB}/2\pi=0.011 \pm 0.003 \, \mu$K$^2$ at $\ell=46$,
a value consistent with the range of estimates presented
in section~\ref{sec:dust}.  The two panels in Fig.~\ref{fig:spectra} compare the best-fit foreground-only
model and the best-fit gravitational wave-only model, both with lensed
$E$-modes added, to the \Bicep\ and \B$\times$\Bicep\ data.  Following the
\Bicep\ analysis, the fit only uses the five lowest multipole bins.
We include cosmic variance errors and use a
foreground-based spectrum for the fiducial model in the
Hamimeche-Lewis likelihood computation.  In addition, we repeat the
likelihood calculation using a Gaussian approximation and find good
agreement.  The null hypothesis predicts an effective spectral index
$\beta \approx 1$ between 100 and 150 GHz.

Fig.~\ref{fig:spectra} suggests that in the absence of a prior the foreground-only model is as good a fit
to the data as the gravitational wave-only model.  To quantify this, we compare
simple Gaussian $\chi^2$ values for the models for a covariance matrix that includes sample variance for a gravitational wave signal with $r=0.2$. 
Using only the five lowest multipole bins of the $150\!\times\!150$ GHz data, we find $\chi^2 = 1.1$ for the best-fit gravitational wave-only
model and $\chi^2 = 1.7$ for the best-fit foreground-only model.  Using all
nine multipole bins in the \Bicep\ $150\!\times\!150$ GHz power spectrum, we find $\chi^2 = 8.5$ for the best-fit gravitational wave-only
model and $\chi^2 = 7.2$ for the best-fit foreground-only model. Thus, in the absence of a prior on the dust contribution, the gravitational wave-only model and the foreground-only model fit the 150 GHz data equally well.

\begin{figure}
\includegraphics[trim=.05cm 0.2cm 0cm 0.cm,width=2.3in]{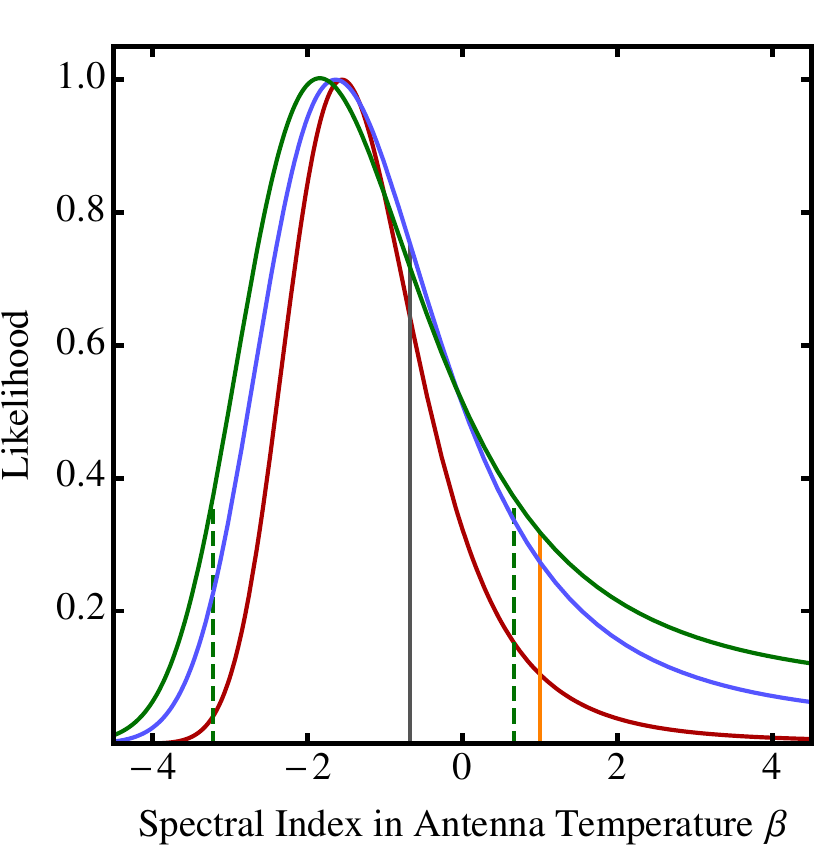}
\caption{Likelihood of the spectral index of the signal 
  in antenna temperature given the $100\!\times\!150$ GHz \B$\times$\Bicep\ and 
  $150\!\times\!150$ GHz \Bicep\ data.
    The red curve shows the posterior for the spectral index of the signal derived from a fiducial Gaussian approximation to the likelihood. As in the \Bicep\ analysis no correction is applied for the lensing contribution. It lies very close to their published result.
  The blue curve uses a likelihood function corrected to account for CMB lensing. In both cases the fiducial model is a $\Lambda$CDM model with $r=0$. The green curve accounts for lensing and uses a Gaussian approximation to the likelihood that includes the variance
  associated with a foreground characterized 
  by an angular power spectrum with $\ell(\ell+1)C_\ell^{BB}/2\pi=0.01\, \mu$K$^2$ at $\ell=46$, 
  $\ell$-dependence consistent with dust, and spectral index $\beta$. 
  The covariance matrix accounts for the correlations between the $100\!\times\!150$ GHz 
  and $150\!\times\!150$ GHz data. Because our analysis accounts for CMB lensing, the likelihood function is broader than that
  computed in~\cite{B2}.  The vertical lines in the plot denote the best-fit CMB prediction
  (black) and the best-fit foreground prediction (orange). The dashed
  line shows the 68\% confidence interval. The null-hypothesis is not convincingly excluded, 
  but a CMB spectrum provides a slightly better fit. However, the constraint on the spectral index is 
  entirely driven by the second bandpower at $100\!\times\!150$ GHz. 
\label{fig:spectral_index}}
\end{figure}

Next, we compute the joint likelihood\footnote{Our covariance matrix includes cosmic variance both in the diagonal and off-diagonal blocks to account for correlations between the $100\!\times\!150$~GHz and $150\!\times\!150$~GHz data. } of the $100\times150$ GHz
\B$\times$\Bicep\ data and $150\!\times\!150$ GHz \Bicep\ data as a
function of the spectral index of the signal. Following the same procedure
as used
in~\cite{B2}, we parameterize the theory input in terms of five bandpowers at $150\times150$ GHz and the spectral index in antenna temperature
$\beta$, and marginalize over the five bandpowers. Note that the
analysis in~\cite{B2} also includes the $100\times100$ GHz \B\ data, which
is not publicly available. Our reproduction of the analysis done
in~\cite{B2} is shown in red in Fig.~\ref{fig:spectral_index}. The
good agreement with the constraint on the spectral index derived
in~\cite{B2} shows that the $100\!\times\!100$ GHz \B\ data does not
significantly impact the constraints on the spectral index.
We also show the results of two modified
analyses: (1) we account for the contribution of the lensed $E$-mode signal
(blue); and (2) we use cosmic variance error bars of the
$\Lambda$CDM model as well as a dust contribution that fits the $150\times150$ GHz data (green), rather than an $r=0$
$\Lambda$CDM model with no foregrounds. The green curve also accounts for the
lensed $E$-mode contribution.  Finally, we indicate the
spectral index predicted for a model with 70\%-correlated polarized
synchrotron and dust emissions in orange.
The final likelihood function (green posterior) is broad enough that
the contribution to the signal not due to lensing is
consistent with either Galactic foregrounds or gravitational waves. 
This is confirmed by a combined fit to the first five bandpowers of both the $100\!\times\!150$ GHz \B$\times$\Bicep\ and
  $150\!\times\!150$ GHz \Bicep\ data. We find $\chi^2 = 8.2$ for the best-fit gravitational wave-only
model and $\chi^2 = 9.9$ for the best-fit foreground-only model. The slightly better fit for the gravitational wave-only scenario is entirely
driven by the $\ell=73$ bandpower in the $100 \times 150$ GHz \B$\times$\Bicep\ data. Figure~\ref{fig:spectral_index} and values of $\chi^2$ rely on a fiducial Gaussian approximation to the likelihood. We have also implemented the Hamimeche-Lewis approximation~\cite{Hamimeche:2008ai} and find good agreement. 

We conclude that the current \B\ and \Bicep\ data cannot distinguish
between the $r=0.2$ model and the null hypothesis based on a spectral
analysis.  However, the upcoming 100~GHz data from the Keck Array
could potentially strongly prefer one of these options.  The null
hypothesis ($r=0$ plus foregrounds) predicts that at $100\times100$~GHz \mbox{$\ell(\ell+1)C_{\ell}^{BB}/2\pi<0.01 \, \mu {\rm K}^2$} 
at $\ell = 100$ even under conservative assumptions about
synchrotron and dust, while if $r =0.2$, the amplitude of the signal
should be significantly larger, $\sim\!0.015 \, \mu {\rm K}^2$ (in the
absence of any foreground emission).  
Moreover, the uncertainties in the synchrotron amplitude on these
scales can be reduced 
through cross-correlations between the \WMAP\ K-band data and the Keck
100 GHz data. Thus,  the Keck 100 GHz measurements may help clarify the
nature of the fluctuations seen by \Bicep\ at 150 GHz.

\section{Estimating the Dust Polarization Signal}
\label{sec:dust}

\begin{figure}
\includegraphics[width=4.in]{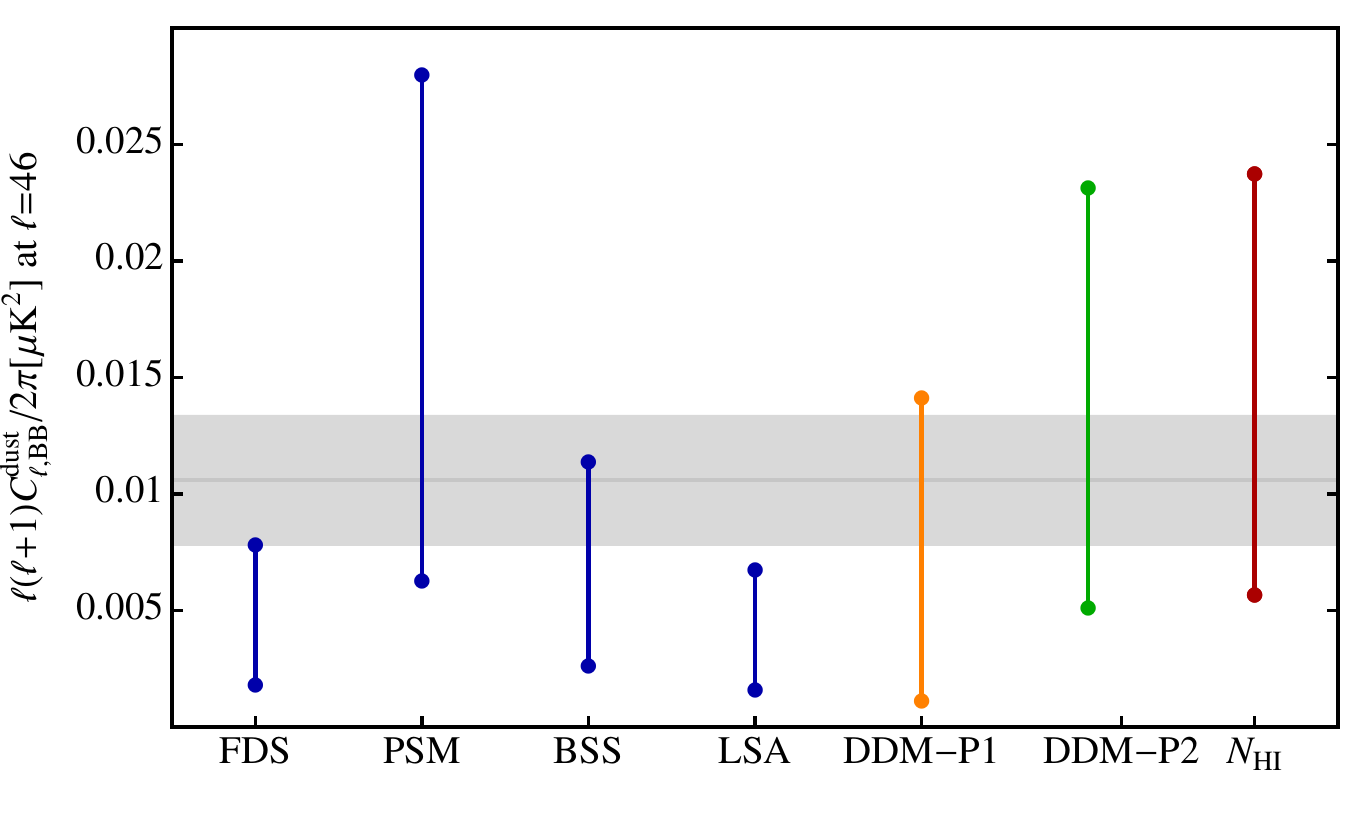}
\caption{Predicted contribution of polarized dust emission 
  to the $B$-mode angular power spectrum for our models discussed in
  section~\ref{sec:dust}, and for the pre-\Planck\ models
  studied by \Bicep\ (blue) after taking into account the increase in
  polarization fraction. The range for the FDS, PSM, BSS, and LSA
  models, shown in blue, is based on a variation of the polarization
  fraction between 8 and 17\%, while the range for the DDM-P1 and
  DDM-P2 models is based on our set of 96 models (see
  section~\ref{sec:dust}). The range for the {\sc Hi} estimate reflects the
  uncertainty in the extrapolation to low column densities and the
  uncertainty in frequency extrapolation. The gray band shows the
  best-fit amplitude of $0.011 \pm 0.003 \, \mu$K$^2$ at $\ell = 46$ determined in section \ref{sec:multi_fit}.
  If the dust foreground amplitude lies in this gray band, then the best-fit model to the data will have a negligible gravitational wave contribution.}
\label{fig:template}
\end{figure}

The \Bicep\ team used the auto-correlations of several dust model templates as well 
as the cross-correlations of these templates with their data 
to model the polarized dust emission in the \Bicep\ region
and to conclude that the polarized dust contribution only modifies the likelihood slightly.  The analysis is based on six dust models: four of the
templates referred to as FDS, BSS, LSA, and PSM are based on pre-\Planck\ data, while the remaining two,
DDM1 and DDM2, are driven by polarization information presented at the
April~2013 ESLAB meeting.  We start with a brief description of these models.

DDM1 uses the \Planck\ map of Galactic thermal dust emission, which is
obtained from fitting \Planck\ 353, 545, and 857 GHz data, as well as
IRAS 100 $\mu$m data~\cite{Planck2013dust}.  This map is constructed
at $353$ GHz and then scaled to $150$ GHz using a modified blackbody
SED with constant emissivity $1.6$ and constant temperature $19.6$ K.
The amplitude of polarized dust emission is then set by assuming a
uniform $5$\% dust polarization fraction over the \Bicep\ field, and
$Q$ and $U$ maps are finally derived using polarization angles from
the Planck Sky Model (PSM)~\cite{Delabrouille2013}.  PSM
predictions are currently not based on \Planck\ data, but rather rely
on modeling informed by earlier experiments.

DDM2 uses the same dust intensity map as DDM1, but relies on a
digitization of the polarization fraction and polarization angle maps
presented in \cite{BernardESLAB} to construct $Q$ and $U$ maps. 

Since DDM1 does not include fluctuations in the polarization fraction, it is expected to under-predict the dust contribution to the power spectrum. For DDM2, noise bias and noise in the polarization fraction map will bias its prediction high.

The \Bicep\ analysis of DDM1 and DDM2 shows polarized dust emission to be subdominant.
However, this conclusion rests on a crucial input, the dust polarization fraction $p$ in the \Bicep\ field, which enters quadratically in the 
dust polarization auto-spectra.

The polarization fraction is also an important parameter for the
remaining four models presented in~\cite{B2}, and was not well
constrained when these models were made. A study dedicated to an
understanding of the role of foregrounds for a potential future space
mission (CMBPol) estimates an uncertainty of one order of magnitude for the polarization fraction~\cite{Dunkley:2008am} and hence as much as two orders of magnitude in the power spectrum. For the FDS model (based on Model 8 in~\cite{Finkbeiner1999}), the \Bicep\ analysis assumes an average polarization fraction in the patch
of $5\%$, in agreement with the average we obtain from the map shown
in~\cite{BernardESLAB}, and sets the Stokes $Q$ and $U$ parameters equal so that $C_\ell^{BB}=C_\ell^{EE}$. The
polarization maps for the BSS and LSA models are also constructed from Model 8 in~\cite{Finkbeiner1999}, with polarization fraction and angles determined from
magnetic field models and line-of-sight integrals. The magnetic field
model for the BSS model is a bi-symmetric spiral, and for LSA a
logarithmic spiral arm. The polarization fraction is normalized to
$3.6\%$ within the  \WMAP\  P06 mask for both models. This yields an
average polarization fraction of $5.7\%$ and $4.9\%$ in the
\Bicep\ region for the BSS and LSA models, respectively, again close
to the polarization fraction obtained
from~\cite{BernardESLAB}. Finally, for the PSM model (based on Model 7 in~\cite{Finkbeiner1999}) we find an
average polarization fraction in the \Bicep\ region of $5.5\%$.\footnote{We run version 1.7.8 of the PSM with the same settings as BICEP to facilitate comparison, {\it i.e.} run as `prediction', with magnetic field pitch angle of $-30^\circ$ and $15\%$ intrinsic polarization fraction. This mode misses some information about scales smaller than 3 degrees and will underpredict the degree-scale power spectrum. Simulated small scale structure is added when run in `simulation' mode.} If the
true polarization fraction were different, all these models would have
to be rescaled. In other words, while only DDM1 and DDM2 
explicitly rely on~\cite{BernardESLAB}, the other four models implicitly
depend on it as well.

The polarization fraction presented in~\cite{BernardESLAB} was not intended 
for quantitative analysis, and there are significant uncertainties in its interpretation.
One important uncertainty is whether the intensity map in the denominator contained a CIB contribution, in particular a CIB monopole.
Visual comparison of the recently published polarization fraction map in~\cite{Planck2014dustoverview} (their
Fig.~4) to that in~\cite{BernardESLAB} shows higher polarization fractions in~\cite{Planck2014dustoverview} than in~\cite{BernardESLAB}. This suggests that a CIB contribution was present in the original map used by \Bicep. 
Based on section 2.4 of~\cite{Planck2014dustoverview}, the CMB may also not have been
subtracted, but this potential correction is small enough to be
negligible ($\sim\!10\%$).  As a result, the polarization fraction $p_{\rm Gal-B2}=\sqrt{Q_{353}^2 + U_{353}^2}/{I_{353}} $
assumed by \Bicep\ underestimates the Galactic dust
polarization fraction $p_{\rm Gal-Actual} =\sqrt{Q_{\rm Gal}^2 + U_{\rm Gal}^2}/I_{\rm Gal} $ by 
\beqn
p_{\rm Gal-B2} & = & \frac{\sqrt{Q_{353}^2 + U_{353}^2}}{I_{353}} \nonumber \\
   & \approx & \frac{\sqrt{Q_{\rm Gal}^2 + U_{\rm Gal}^2}}%
                    {I_{\rm Gal} + I_{\rm CIB} + I_{\rm CMB}} \nonumber \\
         & = & \frac{I_{\rm Gal}}{I_{\rm Gal} + I_{\rm CIB} + I_{\rm CMB}}\, p_{\rm Gal-Actual}\,,
\label{eq.pfrac}
\eeqn
Here $I_{\rm Gal}$ is the Galactic dust intensity and $Q_{\rm Gal}$ and $U_{\rm Gal}$ are the Galactic dust Stokes parameters, all
at 353~GHz, $I_{\rm CIB}$ is the CIB intensity, $I_{\rm CMB}$ is the CMB intensity, and going from the first to the second line in
Eq.~(\ref{eq.pfrac}) assumes that the CMB polarization level is
negligible compared to the dust polarization amplitude at
$353$~GHz~\cite{Planck2014dustoverview}, and that the CIB is
essentially unpolarized. Because the CIB originates from a large
number of galaxies with random polarization orientations, the latter
is a good approximation.  Eq.~(\ref{eq.pfrac}) should be interpreted
as a function of position on the sky $\hat{n}$, and can immediately be
inverted to construct a map of the dust polarization fraction
$p_{\rm Gal}(\hat{n})$ from \cite{BernardESLAB}.
After applying a CIB and CMB correction according to~\eqref{eq.pfrac} to~\cite{BernardESLAB}, using the \Planck\ dust model as template for $I_{\rm Gal}$, we find an average dust polarization
fraction of 16\%,\footnote{The \Planck\ dust model does not contain a CIB monopole, but contains residual CIB fluctuations. These are significant in low emission regions of the sky for $\ell\gtrsim 100$, but are negligible on the large scales probed by the average polarization fraction. Thus, this is an estimate of the dust polarization fraction.} more than a factor of 3 greater than in the uncorrected map.

As a ratio of maps, the polarization fraction map depends on the zero-levels of intensity and polarization maps. Especially in regions of low emission, the uncertainties in these zero-levels translate into significant uncertainties on the polarization fraction. Marginalization over the zero levels in~\cite{BernardESLAB} suggests a polarization fraction of $11^{+6}_{-2}\%$ with a minimum of 8\% and a long tail toward high values. These maps are very preliminary and before the release of the \Planck\ polarization data a detailed analysis can only be done by the \Planck\ collaboration. However, both estimates support polarization fractions a factor $2-3$ larger than those assumed by \Bicep, indicating that the emission from dust in the region is highly polarized. This is consistent with the high polarization fractions seen in synchrotron emission in this region of the sky and these numbers are well within the range of polarization fractions seen at high Galactic latitude~\cite{Planck2014dustoverview}.

 While the average polarization fraction is a good indicator of how large polarized foregrounds are expected to be, it is conceivable that the polarized emission is significantly stronger on large scales than on smaller scales. 
 
All estimates of the dust polarization power spectrum based on auto-correlations scale as the square of the polarization fraction. Motivated by refs.~\cite{BernardESLAB} and~\cite{BoulangerESLAB}, we consider a range of $8 - 17$\%, which leads to an increase of all estimates in~\cite{B2}, including those based on pre-\Planck\ templates, by at least a factor $2.5$ and as much as an order of magnitude.
 The revised estimates are shown in Figure~\ref{fig:template}.
For the pre-Planck models, the blue bands in the plot reflect the uncertainty in the polarization fraction arising from its sensitivity to zero-levels in the polarization and intensity maps. We see that the contribution of the dust polarization according to the models is comparable to or perhaps even larger than both the \Bicep\ data and the best-fit amplitude of $0.011 \pm 0.003 \, \mu$K$^2$ at $\ell=46$ estimated in section~\ref{sec:multi_fit}, which is shown as the gray band. 
Thus, a more detailed investigation is necessary, and we present ours in what follows.

\subsection{Revised Data-Driven Models}
\label{sec:DDM}

Using the CIB-corrected polarization fraction map, we construct our
own 150~GHz data-driven dust models (\mbox{DDM-P1} and DDM-P2) and
compute their auto-correlations.  We scale the 353~GHz intensity map
to 150~GHz using the scaling recently reported
in~\cite{Planck2014dustfreq}, and assume that the polarization
fraction does not change significantly between these two frequencies.
Although \cite{Planck2014dustfreq} reports a decrease in the
polarization fraction at 150 GHz relative to 353 GHz, this effect is
small and does not affect our conclusions.  

DDM-P1 uses the average polarization fraction across the region. This ignores polarization fraction fluctuations, biasing the predicted dust polarization levels low. Templates such as the CMB-free 353~GHz map or even the \Planck\ dust model contain residual CIB fluctuations. These have a negligible effect on the average polarization fraction, but lead to an overestimate of the dust polarization power spectrum on small scales ($\ell\gtrsim 100$) in DDM-P1 or any other model that relies on average polarization fractions. For DDM-P1, we thus estimate the polarized emission from dust at $\ell=46$ and assume $C_\ell\propto \ell^{-2.4}$. 
DDM-P2 uses the spatially varying polarization fraction from the CIB-corrected map and avoids the subtleties inherent to models that rely on average polarization fractions such as DDM-P1. However, noise bias and noise in the polarization fraction map are expected to cause DDM-P2 to slightly over-predict the polarized emission from dust. 

For both DDM-P1 and DDM-P2 we consider models with different estimates for the polarization angles. 
Interstellar dust grains preferentially absorb the optical light from
stars in the direction perpendicular to the Galactic magnetic
field. As a result, their emission is polarized in this same
direction, \ie\ perpendicular to the starlight polarization
direction~\cite{DraineBook}.  We gathered data in the \Bicep\ region
from the Heiles, Santos, and Schr\"oder
samples~\cite{Heiles,Santos,Schroder}.  Since this region has been
selected by the \Bicep\ team for its low dust extinction, few
starlight polarization data have been collected within the field.
However, we found seven significant detections ($P/\sigma_P>1$) along
sightlines to stars at least 100~pc above the Galactic plane.  Two of
them are for the same star, but observed by different teams, with both observations above
5$\sigma$.  The polarization angle of the dust emission derived from
the latter is $154.5^\circ$.  The mean and median angles derived from
all significant detections in the region are respectively
$171.1^\circ$ and $160.4^\circ$, in good agreement with that derived
from the $5\sigma$ detections.  
In a first class of models, we thus take the polarization angle to be constant across the patch, and explore a range of values consistent with starlight polarization data, taking the average dust emission polarization angle to be $160^\circ$, and explore the effect of varying this angle by $10^\circ$.  

In a second class of models, we again take the polarization angle to
be constant across the patch, but use the average polarization angle
from the PSM. We consider a third class of models, in which we use polarization angles derived from the PSM after smoothing the maps to 1 or 5 degrees. Finally, we consider models based on~\cite{BoulangerESLAB} and vary the zero levels of the polarization and intensity maps within errors of the calibration.

\begin{figure}
\includegraphics[width=2.3in]{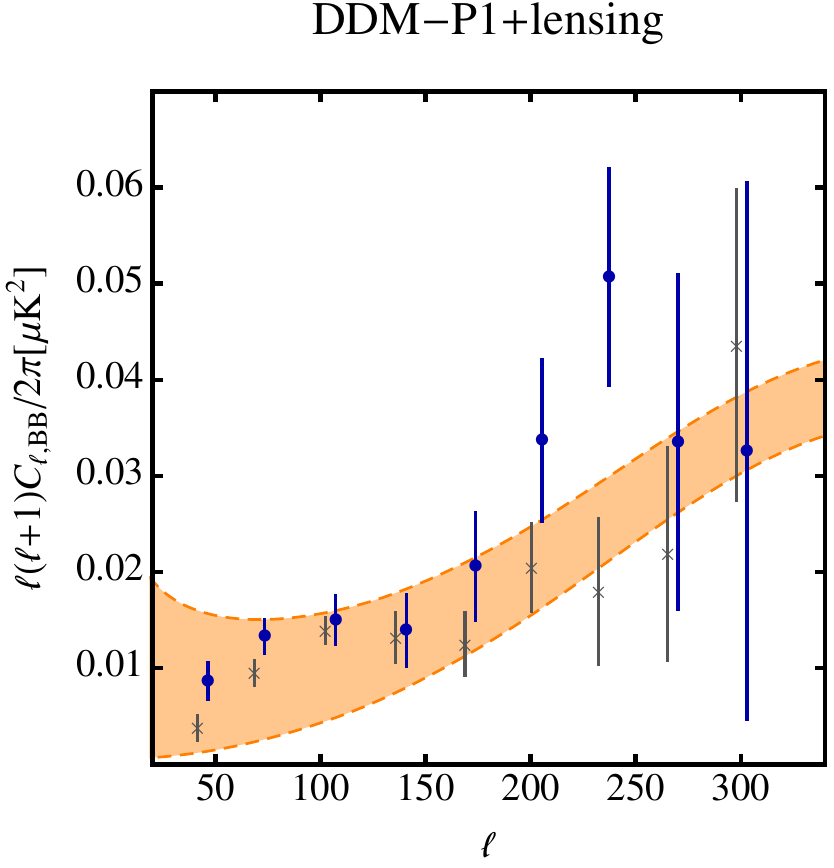}
\includegraphics[width=2.3in]{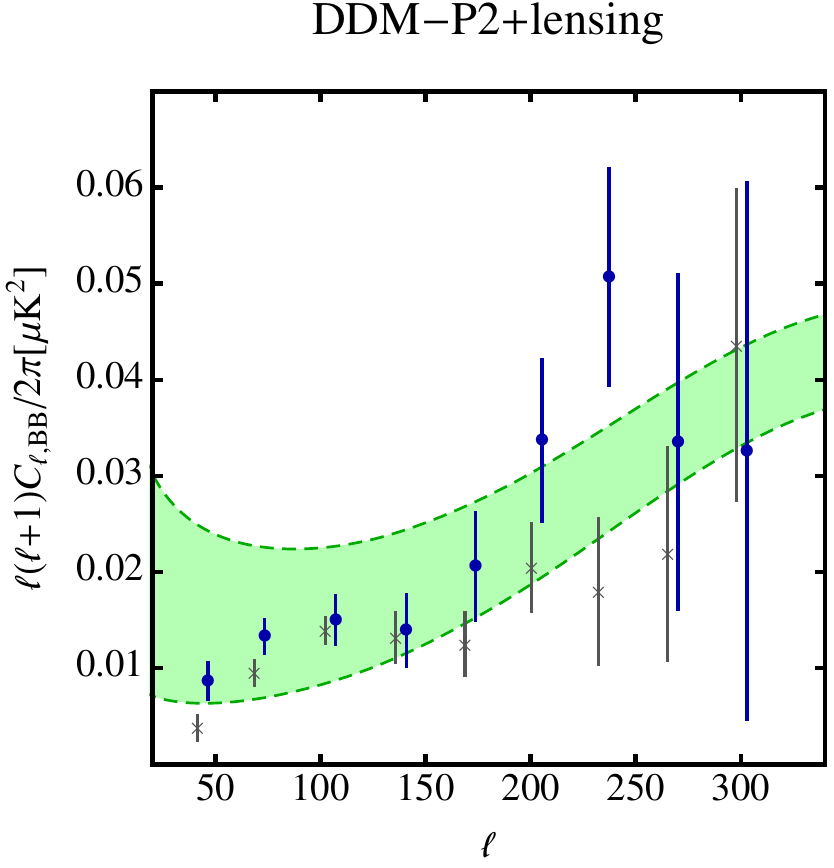}
\includegraphics[width=2.3in]{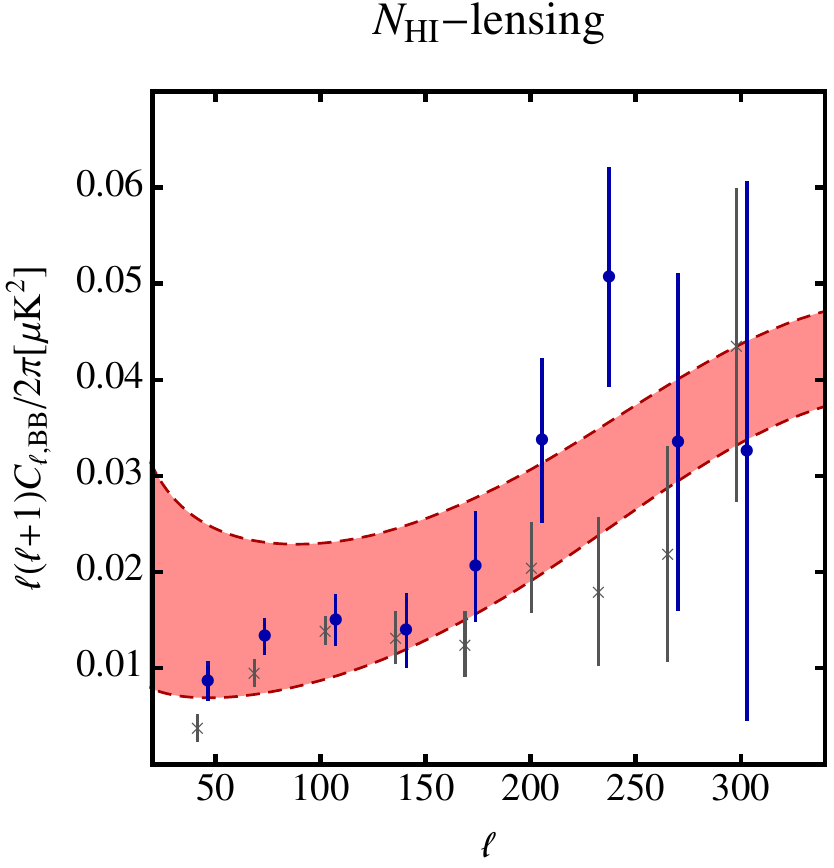}
\caption{Comparison of several predictions for the 150~GHz signal versus
  the reported \Bicep\ $\times$ \Bicep\  and  the preliminary \mbox{\Bicep\ $\times$ Keck} measurements. 
The predictions are a combination of the  dust polarization signal and the predicted lensing signal for standard cosmological parameters.
Panel (a) is based on DDM-P1, which assumes that the dust polarization signal is proportional to the dust intensity (extrapolated from 353 GHz)
times the mean polarization fraction (based on our CIB-corrected map;
see section~\ref{sec:dust}). The band represents the 1$\sigma$
contours derived from a set of 48 DDM-P1 models.  Panel (b) shows
DDM-P2, with polarization fractions from our CIB-corrected map, and
polarization direction based on starlight measurements, the PSM,
or~\cite{BoulangerESLAB}.  Panel (c) uses the column density of
neutral hydrogen in the \Bicep\ region inferred from the optical depth
at 353 GHz to estimate the dust foreground.  In this panel, the band reflects the uncertainty in the extrapolation of the scaling relation to low column densities as well as the uncertainty in the rescaling from 353 GHz to 150 GHz. \label{fig:power_spec}}
\end{figure}

The first two panels of Fig.~\ref{fig:power_spec} show the range of
dust $B$-mode amplitudes compatible with each model added to the
lensed $E$-mode signal.  The DDM-P1/DDM-P2 envelopes correspond to the
1$\sigma$ contours based on a suite of forty-eight DDM-P1/DDM-P2 models that
differ by their choice of polarization angles and map zero-levels, as
discussed above.
DDM-P1 and DDM-P2 lead to consistent predictions, and the uncertainty
envelope on each estimate encompasses the \Bicep\ and
\Bicep\ $\times$ Keck data points in the five bins used in the
\Bicep\ analysis.

\subsection{Estimate from {\sc Hi} Column Density}
\label{sec:HI}

The \Planck\ collaboration has reported a strong correlation between
{\sc Hi} column density and the amplitude of the dust polarization
signal along a given line of sight~\cite{AumontESLAB}.  We use this
relationship to estimate the polarization signal in the
\Bicep\ region.  {\sc Hi} column density can be inferred from the
\Planck\ $353$~GHz dust opacity map according to
$N_{\text{\sc Hi}}=1.41\times10^{26}\,{\rm cm}^{-2}\,\tau_{353}$~\cite{Planck2013dust}. 
Using this relation, we find
$ N_{\text{\sc Hi}} = (1.50 \pm 0.07) \times 10^{20}\, {\rm cm}^{-2}$ in the
\Bicep\ region.\footnote{While Ref.~\cite{AumontESLAB} was based on an
  older version of the \Planck\ dust model, we consistently work with
  version 1.20.}  
Inserting this value into the relation between $N_{\text{\sc Hi}}$ and dust
polarization amplitude
and using the appropriate modified blackbody SED~\cite{Planck2014dustfreq},
at $150$~GHz we obtain polarized dust emission power estimates at $\ell=100$ of
\mbox{$\ell(\ell+1)C_{\ell}^{EE}/2\pi=0.021 \pm 0.014 \,\mu{\rm K}^2$} 
and $\ell(\ell+1)C_{\ell}^{BB}/2\pi=0.015 \pm 0.010 \,\mu{\rm K}^2$.

The third panel in Fig.~\ref{fig:power_spec} shows the predicted range
of the polarized Galactic dust emission inferred from the above
procedure combined with the lensed CMB signal.  The primary caveats of
this method are that it involves applying scaling relations calibrated
for large sky fractions and high {\sc Hi} column densities to the \Bicep\ region, which covers only
$\sim\!1$\% of the sky and has a low {\sc Hi} column density, and that it assumes that the same scaling
relations hold at $353$~GHz and $150$~GHz.  In addition, the scatter
in the relations is likely to increase as one proceeds to smaller 
{\sc Hi} column densities similar to those in the \Bicep\ region.  With
these caveats in mind, we note the good agreement between the
{\sc Hi} estimate and those relying on the DDM-P1 and DMM-P2 models
shown in the leftmost panels of Fig.~\ref{fig:power_spec}.
This suggests that although each of the methods used to derive the estimates shown in Fig.~\ref{fig:power_spec} comes with its own caveats, the prediction of a high level of 150~GHz polarized dust emission in the \Bicep\ field is robust.

\subsection{Consistency Check}
\label{sec:direct}

We compare our estimates for the dust polarization signal in the
\Bicep\ region to a measurement of the angular power spectrum from the
maps presented in~\cite{BoulangerESLAB} in the \Bicep\ region. To
understand the properties of these maps, we have reproduced the
analyses in~\cite{AumontESLAB} and find good agreement. Furthermore,
we have computed cross-spectra with the \WMAP\ W-band data and find
good agreement with the 100$\times$353~GHz spectra presented
in~\cite{AumontESLAB}. Additional cross-correlations with lower
frequency \WMAP\ data provide further evidence that the maps are
reliable, making a measurement of their power spectrum worthwhile.  We
stress that none of the conclusions in this paper rely on digitizing
the maps in~\cite{BoulangerESLAB}.  However, this exercise constitutes
a useful cross-check of the work already presented, and also allows us
to demonstrate the internal consistency of the results
presented by
\Planck\ in~\cite{BernardESLAB,BoulangerESLAB,AumontESLAB}.

Our power spectrum estimator is based on PolSpice~\cite{PolSpice}.  We
have carefully set the PolSpice parameters to ensure that we recover
the input angular power spectra on the \Bicep\ patch using 250 
CMB-only simulations, filtered with the \B\ filter function. Our filtering is isotropic, while that of \Bicep\ is not. However, given the high degree of isotropy of the filtered dust maps we do not expect this to alter the conclusions. 

The maps contain two sources of noise that are a
priori unknown: the noise of the instrument, and the noise introduced
by the digitization. For the instrumental noise, we assume that its
shape is well described by the \Planck\ noise model with parameters
identical to those for intensity. To understand the noise introduced
by digitization, we have developed a pipeline that takes HEALPix maps,
converts them to GIF files, and inserts them into a presentation which
is then saved as a PDF file. We then apply our digitization procedure
to convert the PDF files back to GIFs and then to HEALPix data
files. At 353 GHz, the polarized emission is dominated by dust. We
thus apply this pipeline to ten simulations of dust maps. This has
allowed us to characterize the effects introduced by the digitization
procedure in the form of a transfer function. In addition to the dust,
the maps contain instrument noise. We thus process ten noise
simulations through our pipeline and measure the corresponding
transfer function. We assume that the power spectrum of the map is
well described by $C^\text{obs}_\ell =
F_\ell^\text{Dust}C_\ell^\text{Dust} + A F_\ell^\text{Noise}N_\ell$,
where $F_\ell$ are the transfer functions measured in the simulations,
$C_\ell^\text{Dust}$ is the underlying dust power spectrum, $N_\ell$
is the \Planck\ noise model with parameters extracted from a fit to
the half-ring half-difference for the 353 GHz intensity maps, and $A$
is a fitted~amplitude.

We can extrapolate our measurement at 353 GHz to 150 GHz using the dust properties from~\cite{Planck2014dustfreq} as before to predict
the dust polarization signal. The resulting level is consistent with
the predictions from the previous methods shown in
Fig.~\ref{fig:power_spec}.  Given that the maps in~\cite{BernardESLAB}
have not yet been released by \Planck, we choose not to show the
derived spectrum next to those in Fig.~\ref{fig:power_spec} as an
acknowledgment of the preliminary nature of these maps.  We only
consider them here as a final cross-check, as even in their
preliminary form, they represent the best publicly available maps of
dust polarization, as well as a significant improvement over our
pre-\Planck\ knowledge of foregrounds.


\section{Cross-Correlation Estimates from Polarization Templates}
\label{sec:x-corr}

\begin{figure}[t]
\includegraphics[trim=1.2cm 1.3cm 1.3cm 1.3cm,clip,width=3in]{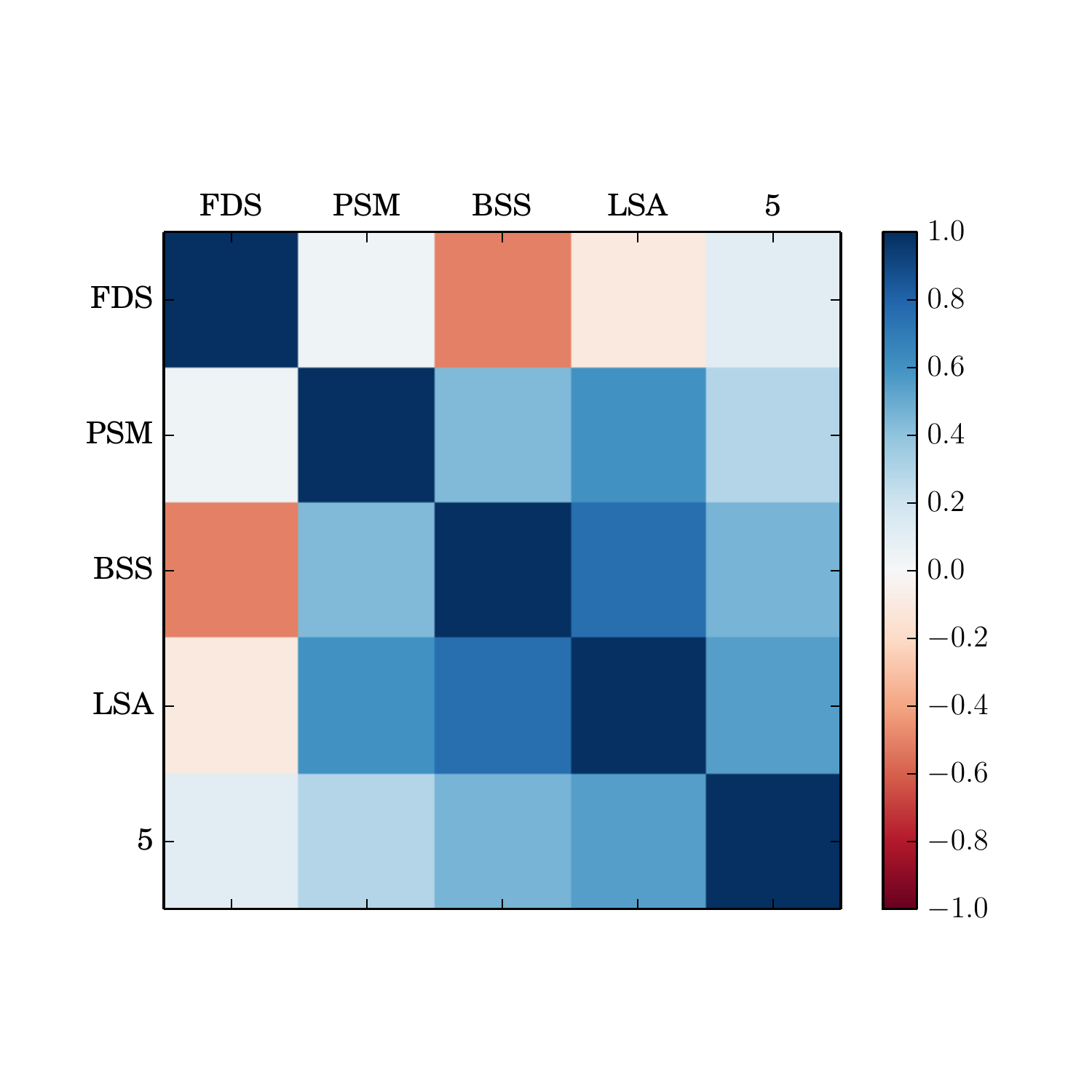}
\includegraphics[trim=1.2cm 1.3cm 1.3cm 1.3cm,clip,width=3in]{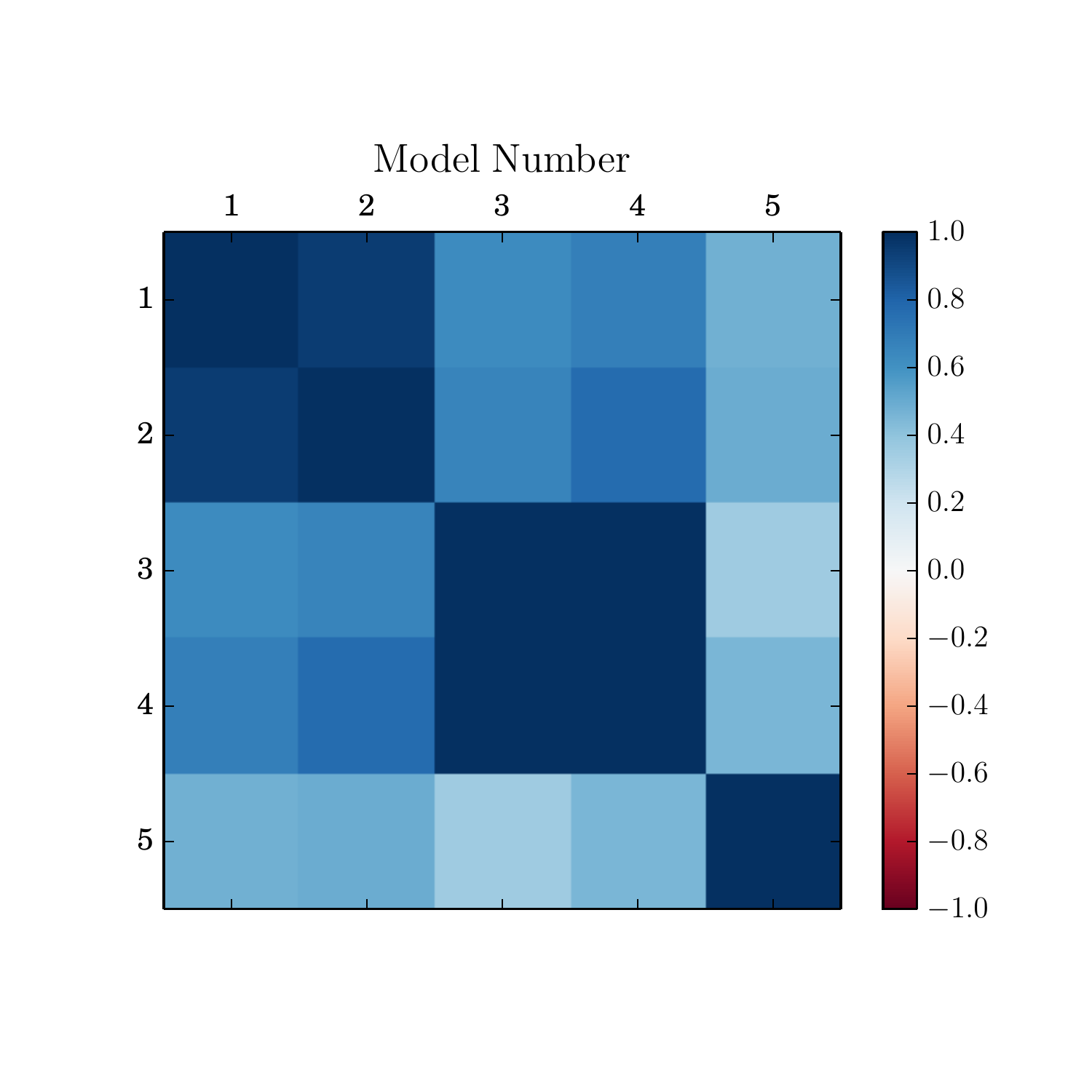}
\caption{The left panel shows the correlation matrix at $\ell=46$ for model 5 and four of the templates
used in~\cite{B2}: the Planck Sky Model (PSM)~\cite{Delabrouille2013},
the Bi-Symmetric Spiral (BSS) and Logarithmic Spiral Arm (LSA) field models presented in~\cite{ODea2012},
and Model 8 of~\cite{Finkbeiner1999} with $Q=U$. If the true sky looked like one of the models,
then a measurement of the cross-correlation using another model would underestimate the signal
by as much as a factor of 10. The correlations further decrease for higher multipoles.
The right panel shows the correlation matrix at $\ell=46$ for a small subset of five DDM-P2 dust
models. The polarization angles are taken to be (1) the average angle
in the patch as inferred from starlight data; (2) the average angle
taken from the PSM; (3) from the PSM at 5 degree resolution; and (4)
from the
PSM at 1 degree resolution. Model 5 is based on~\cite{BoulangerESLAB}
and is a proxy for data. Even between ``data-based"  models and
data, correlation coefficients below 50\% are common, suggesting that low cross-power
between the data-driven models and data do not establish that foregrounds are
negligible.}
\label{fig:corr}
\end{figure}

Cross-correlations with templates that very accurately trace the foreground polarization
can provide a precise estimate of the contribution of the foreground to a map.
However, this requires that there be little noise in the foreground template and that it correctly capture
the spatial structure of the actual foreground.   
If the foreground template differs in spatial structure
from the actual foreground, then any measurement of the cross-correlation will underestimate the contribution of the foreground to the power spectrum.

Since the \Bicep\ maps are not publicly available, we cannot directly test cross-correlations with these maps. However, we can test the templates by measuring their cross-correlations.
If the cross-correlations between the templates are significantly below 1 (as shown in Figure \ref{fig:corr}), then negligible correlations between data and these templates do not imply that foregrounds are negligible.

As discussed in section~\ref{sec:dust}, Ref.~\cite{B2} used a series of templates
that are based on multiplying the intensity of the dust signal by a polarization fraction and a polarization direction.
While the publicly available \Planck\ 353 GHz maps and dust models provide an accurate map of the dust intensity signal,
the polarization directions and the polarization amplitudes are poorly known.  We can estimate the sensitivity of
the measured cross-correlations to the polarization angle by cross-correlating the four publicly available templates used in~\cite{B2} with themselves and the maps from~\cite{BoulangerESLAB}, which we will refer to as model 5 below.		
The matrix of correlation coefficients at $\ell=46$ is shown in the left panel of Figure~\ref{fig:corr}.
The small correlations between the templates and between the templates and model 5 suggest that the small cross-correlations with the data measured in~\cite{B2} likely reflect the limitations of the templates and do not provide a constraint on the dust~polarization.

As a test for the revised data-driven models, we have
computed the full set of cross-spectra for our suite of ninety-six
models (see section~\ref{sec:DDM}).  For clarity, we focus the discussion on a small representative subset of five
DDM-P2 models selected to illustrate the main
conclusions from our analysis.   In all five DDM-P2 models shown, the dust polarization fraction is set from
our CIB-corrected map (see section~\ref{sec:dust}).  We then set
polarization angles in one of five ways: model 1 assumes a constant
polarization angle set from starlight polarization data (see
section~\ref{sec:DDM}); model 2 assumes a constant polarization angle
set from the Planck sky model; models 3 and~4 use spatially-varying
polarization angles again set from PSM maps, but smoothed to 5$^\circ$
and 1$^\circ$, respectively, before computation of the polarization
angles; and model 5 is based on the maps discussed in
section~\ref{sec:direct}.

The correlation matrix for these models at $\ell=46$, corrected for noise bias, is
shown in the right panel of Fig.~\ref{fig:corr}. The two models with constant
polarization angles (models 1 and 2) correlate well with each other,
which is expected since the polarization angles obtained from
starlight data and from the PSM are in good agreement. Similarly, the
models whose polarization angles are based on the smoothed PSM maps
(models 3 and 4) also correlate well with each other, and the
correlations between these models and the first two are still
significant.  However, the correlations between model 5, our proxy for data, and any other
model are typically suppressed by a factor of two to three. Such a
suppression is quite typical in our full $96\times96$ correlation
matrix, and in fact much lower correlation coefficients and even small
negative ones~exist. The correlation coefficients decrease further on smaller scales. 

Model 5 is the only model whose polarization angles are set from
polarized dust emission data.  Although preliminary, as discussed in
section~\ref{sec:direct}, these data are the only of their kind
currently publicly available.  The suppressed correlation between
model~5 and the other models, and more generally between models with
polarization angles set from data and other models, therefore suggests
that the cross-spectra between the six templates studied by
\Bicep\ and their data could significantly underestimate the true foreground level
in their field.  Uncertainties in the spatial
variation of the polarization fraction, turbulence-driven variations in
the polarization direction, and noise in the maps used to generate the template will further suppress the
cross-correlation and lead to even more severe underestimates.  We note that these effects are even larger on smaller
scales and may explain the trends seen in the cross-correlation between DDM2 and the \Bicep\ 150 GHz data
shown in Fig.~6 of~\cite{B2}.

The \Planck\ 217 and 353 GHz data should provide an excellent template
for cross-correlation analysis.  Ref.~\cite{Planck2014dustfreq}
showed that  the polarized dust emission at 150~GHz is expected to be highly (but not perfectly)
correlated with dust at higher frequencies and the noise properties of these maps (when released) should
be well characterized. From the public intensity maps we already know that, because of the \Planck\ scan strategy, the noise levels in the maps in the \Bicep\ region of the sky are about $80\%$ of the average.

In conclusion, cross-spectra with templates that are not based directly on
observed dust polarization data cannot convincingly establish that
foregrounds in the \Bicep\ region are truly negligible.  This
uncertainty will be significantly reduced when the \Bicep\ 150~GHz
maps can be directly cross-correlated with the \Planck\  217 and 353~GHz polarization data.


\section{Conclusion and Outlook}
\label{sec:outlook}

Motivated by the importance of the detection of gravitational waves for our
understanding of cosmology, we have examined the uncertainties in the
amplitude of the dust polarization signal in the \Bicep\ region. 
We conclude that the predicted level of polarized emission from interstellar dust in this field might leave room for a primordial gravitational wave contribution, but could also be high enough to explain the observed excess B-mode power. Thus, no strong cosmological inference can be drawn at this time.

We are in the fortunate situation that the Keck Array 100 GHz maps, the
\WMAP\ K-band and \Planck\ LFI maps, the \Planck\ HFI polarization maps, and the \Bicep\ 150 GHz maps will soon help us
determine the relative contributions of dust, synchrotron, and the CMB to
the signal detected by \Bicep, and may then lead to a definitive discovery of gravitational waves.


\begin{acknowledgments}
We thank Steve Choi, Ren\'{e}e Hlo\v{z}ek, William Jones, Lyman Page, Uros Seljak, Suzanne
Staggs, Paul Steinhardt, and Matias Zaldarriaga for stimulating conversations, Andrei
Berdyugin for pointing us to the Schr\"oder and Santos starlight
polarization catalogs, and in particular Aur\'{e}lien Fraisse for his numerous contributions to the paper. 

RF gratefully acknowledges the Raymond and Beverly Sackler Foundation for their support. RF is also supported in part by NSF grants PHY-1213563 and PHY-0645435.
JCH and DNS are supported by NASA Theory Grant NNX12AG72G and NSF AST-1311756.  
\end{acknowledgments}


\end{document}